\documentclass[modern]{aastex631}
\usepackage{times}

\newcommand{\sdo}{{\it SDO}}%
\newcommand{\hmi}{{\it SDO}/HMI}%
\newcommand{\aia}{{\it SDO}/AIA}%
\newcommand{\stereo}{{\it STEREO}}%
\newcommand{\euvi}{{\it STEREO}/EUVI}%

\shorttitle{Far-side Magnetic-flux Maps}
\shortauthors{Chen et al.}

\begin{document}
\title{Inferring Maps of the Sun's Far-side Unsigned Magnetic Flux from Far-side  Helioseismic Images using Machine Learning Techniques}

\author[0000-0002-2632-130X]{Ruizhu Chen}
\affiliation{W. W. Hansen Experimental Physics Laboratory, Stanford University, Stanford, CA 94305-4085, USA}

\author[0000-0002-6308-872X]{Junwei Zhao}
\affiliation{W. W. Hansen Experimental Physics Laboratory, Stanford University, Stanford, CA 94305-4085, USA}

\author[0000-0002-3631-6491]{Shea Hess Webber}
\affiliation{W. W. Hansen Experimental Physics Laboratory, Stanford University, Stanford, CA 94305-4085, USA}

\author[0000-0002-0671-689X]{Yang Liu}
\affiliation{W. W. Hansen Experimental Physics Laboratory, Stanford University, Stanford, CA 94305-4085, USA}

\author[0000-0001-9130-7312]{J. Todd Hoeksema}
\affiliation{W. W. Hansen Experimental Physics Laboratory, Stanford University, Stanford, CA 94305-4085, USA}

\author[0000-0002-6338-0691]{Marc L. DeRosa}
\affiliation{Lockheed Martin Solar and Astrophysics Lab (LMSAL), Palo Alto, CA 94304, USA}

\begin{abstract}

Accurate modeling of the Sun's coronal magnetic field and solar wind structures require inputs of the solar global magnetic field, including both the near and far sides, but the Sun's far-side magnetic field cannot be directly observed.
However, the Sun's far-side active regions are routinely monitored by helioseismic imaging methods, which only require continuous near-side observations.
It is therefore { both feasible and} useful to estimate the far-side magnetic-flux maps using the far-side helioseismic images despite their relatively low spatial resolution and large uncertainties.
In this work, we train two machine-learning models to achieve this goal. 
The first machine-learning training pairs simultaneous \hmi-observed magnetic-flux maps and \aia-observed EUV 304\,\AA\ images, and the resulting model can convert 304\,\AA\ images into magnetic-flux maps. 
This model is then applied on the \euvi-observed far-side 304\,\AA\ images, available for about 4.3 years, for the far-side magnetic-flux maps.
These EUV-converted magnetic-flux maps are then paired with simultaneous far-side helioseismic images for a second machine-learning training, and the resulting model can convert far-side helioseismic images into magnetic-flux maps.
These helioseismically derived far-side magnetic-flux maps, despite their limitations in spatial resolution and accuracy, can be routinely available on a daily basis, providing useful magnetic information on the Sun's far side using only the near-side observations.

\end{abstract}


\section{Introduction}
\label{sec1} 

While solar flares and coronal mass ejections generally have more impacts on space weather, solar wind structures continually impact the heliosphere, affecting near-Earth conditions and influencing the propagation of intense solar events. 
Prediction of solar wind conditions is therefore a critical aspect of space weather forecasting. 
However, solar wind models are highly dependent on the global magnetic field on the solar surface as their inner boundary condition, and the lack of observations of the Sun's far-side magnetic field poses a significant problem to high-quality space weather forecasting \citep{nit08, sch11}. 
Some global field modelers rely on unupdated near-side observations for the far-side magnetic field, while others estimate the far-side field through flux-transport models \citep[e.g.,][]{sch03, upt14}.
However, active regions (ARs) that emerge on the near side may grow rather than decay after rotating onto the far side, and new ARs may emerge and develop rapidly on the far side of the Sun. 
These behaviors cannot be modeled by the flux-transport models, causing inaccuracies in the far-side magnetic field data currently used by some global coronal-field and solar wind models. 
It has been demonstrated by some authors that including far-side magnetic field with updated flux is important to improved modeling of the background solar wind and the solar wind structure. 
For example, \citet{arg13} showed that the modeled solar wind was in better agreement with observations by plugging an AR, which emerged on the far side and was detected by a helioseismic far-side imaging method \citep{lin00}, in their ADAPT \citep[Air-Force Data Assimilative Photospheric Flux Transport model;][]{arg10} code.

For obvious reasons, observations of the Sun's far side are scarce. 
Carrying a suite of instruments, both spacecraft of the {\it Solar Terrestrial Relations Observatory} \citep[\stereo;][]{how08} observed the Sun's far side for a few years in extreme ultraviolet (EUV) along with other observables, but without magnetic fields. 
Now with one spacecraft having lost communications and one approaching the earth, a large fraction of the Sun's far side is out of our monitoring scope. 
On the other hand, local helioseismology techniques were developed to monitor large ARs on the Sun's far side. \citet{lin00} first reported detection of an AR near the Sun's far-side central meridian using a technique called helioseismic holography \citep{lin00b}, and this imaging technique was further improved in later endeavors \citep{bra01, lin17}.
Using time--distance helioseismology technique, \citet{zha07} developed a new far-side imaging code that included both 4- and 5-skip $p$-mode waves, and their method was improved as well through follow-up studies \citep{har08,ilo09}. 
More recently, \citet{zha19} overhauled the time--distance far-side imaging technique by including more multi-skip waves, which resulted in substantial improvements in their reliability of imaging the Sun's far-side ARs.
It would be interesting to explore the connections of the far-side helioseismic images with the EUV images as well as magnetic flux on the far side, and the knowledge learned from such connections can then be used to map the Sun's far-side magnetic flux by use of the near-side continuous photospheric observations only.

Actually, the 304\,\AA\ observations by \euvi\ \citep[Extreme Ultraviolet Imager;][]{wul04} have already been used to assess the reliability of the helioseismic far-side images. 
In a series of articles, \citet{lie12, lie14, lie17} compared \euvi\ 304\,\AA\ images, in which ARs are seen as regions with enhanced emission, with helioseismic holography far-side images, and found that 95\% of the helioseismically-detected far-side ARs correspond to an EUV brightening area.
Similarly, \citet{zha19} reported that 97.3\% of their time--distance helioseismology-detected far-side ARs that exceed a certain size threshold correspond to a \stereo-observed EUV brightening area.
\euvi\ observations were also used to improve the reliability of helioseismic far-side AR detection.
Recently, \citet{fel19} and \citet{bro21} trained a machine-learning model that compared the far-side EUV Images and the helioseismic-holography images before these helioseismic images were routinely overlapped for 5 days to enhance the signal-to-noise ratio, and found that the trained model was able to detect with a higher confidence smaller far-side ARs, which would otherwise be unrecognized in the routinely-generated far-side helioseismic images.

Neither the \euvi\ observations nor the helioseismic images offer direct magnetic-field information on the far-side ARs; however, past research seemed to suggest that both the 304\,\AA\ images and the helioseismic images were related to magnetic-field strength, or magnetic flux density, of the ARs. 
\citet{uga15} monitored evolution of total \hmi\ \citep[{\it Solar Dynamics Observatory} / Helioseismic and Magnetic Imager;][]{sch12a, sch12b} magnetic flux and total \aia\ \citep[\sdo/Atmospheric Imaging Assembly;][]{lem12} 304\,\AA\ flux for 10 selected ARs, and found that these two quantities have a power-law relation, indicating that the 304\,\AA\ images can be used as a proxy of magnetic flux. 
\citet{kim19} trained a deep-learning code utilizing the \hmi\ magnetograms and \aia\ 304\,\AA\ images, and applied the trained code on the \euvi\ 304\,\AA\ images to derive the Sun's far-side magnetic-field maps.
For far-side helioseismic images, through comparing the helioseismic phase shifts measured in large and stable far-side ARs with the magnetic-field strengths before and after these ARs' far-side passages, \citet{gon07} demonstrated that the far-side ARs' helioseismic phase shifts, which are the physical quantities displayed in the helioseismic images, are highly correlated with their magnetic strengths. 

Based on the substantial progress made by various authors and introduced above, we therefore believe that it is scientifically useful to map the far-side magnetic flux using only the near-side observations, and that it is also practically feasible to do so by combining available observations of near-side magnetic fields, near- and far-side EUV 304\,\AA\ images, and far-side helioseismic images through training two machine-learning models. 
In this work, we train two separate machine-learning models: one model converts EUV 304\,\AA\ images into magnetic-flux maps by coupling the near-side \hmi\ magnetic-field observations and \aia\ 304\,\AA\ images; and the other model converts helioseismic far-side images into far-side magnetic-flux maps, with the \euvi\ 304\,\AA\ images as a bridge between these two sets of conversions. 
This article is organized as follows: we introduce our analysis procedure and data preparation in Section~\ref{sec2}, introduce our first machine-learning code training in Section~\ref{sec3}, and the second machine-learning code training in Section~\ref{sec4}. 
We then summarize and discuss our results in Section~\ref{sec5}.

\section{Analysis Procedure and Data Preparation}
\label{sec2}

\subsection{Analysis Procedure}
\label{sec21}

\begin{figure}[!ht]
\centering
\includegraphics[width=0.7\textwidth]{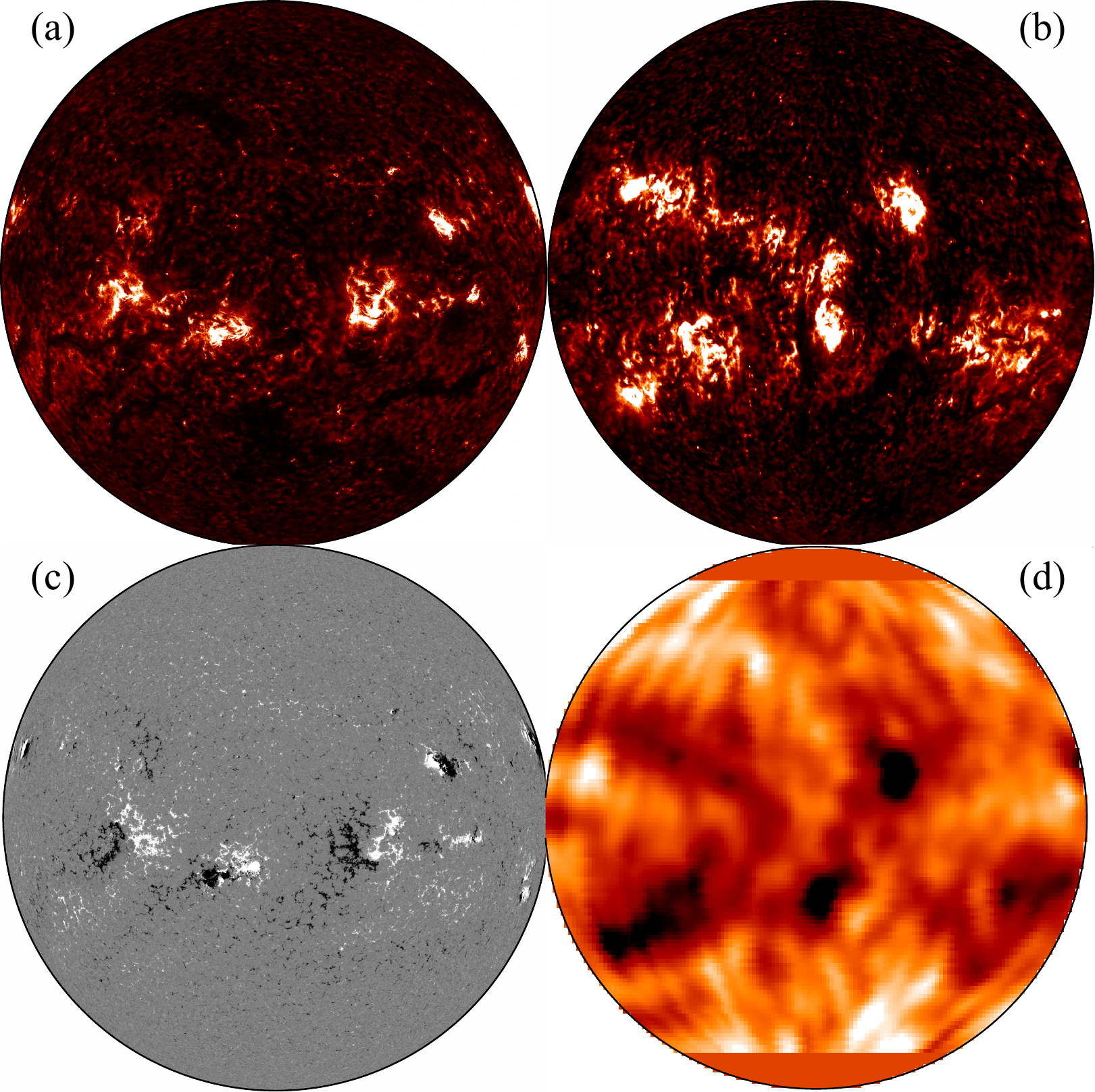}
\caption{Comparison of near-side and far-side images taken at or near 00:00 UT of 2014 March 13. 
(a) \aia-observed near-side 304\,\AA\ image; 
(b) {\stereo}/EUVI-observed far-side 304\,\AA\ image after merging observations from both {\stereo}-A and {\stereo}-B spacecraft. 
(c) \hmi-observed near-side magnetic field; and
(d) time--distance helioseismic far-side image obtained from the \hmi-observed near-side Dopplergrams. 
The magnetic field is displayed with a scale of $-$200 to 200 Gauss, the helioseismic image is displayed with a scale of phase shifts from $-$0.10 to 0.10, and the EUV images are displayed with a DN from 10 to 2000. 
The \stereo/EUVI 304\,\AA\ image is displayed after its intensity is calibrated to match the \aia\ 304\,\AA\ level.}
\label{composite_img}
\end{figure}

As demonstrated by \citet{uga15}, practiced by \citet{kim19}, and clearly seen through comparing Figure~\ref{composite_img}a and \ref{composite_img}c, the Sun's magnetic-field strength and its EUV emission, particularly in the wavelength channel of 304\,\AA, are well correlated, and one quantity can be converted to the other, likely with a loss of accuracy. 
However, when one quantity is unavailable, e.g., in this case, the far-side magnetic flux density, a conversion with some losses of accuracy is still an acceptable approximation, as long as its limitation is recognized and its errors are well estimated. 
Similar to what \citet{kim19} did, in this work we can train our first machine-learning model using simultaneous observations of \hmi\ magnetic-field strengths and \aia\ 304\,\AA\ images, and then apply the trained model on the recalibrated \euvi-observed far-side 304\,\AA\ images for the Sun's far-side magnetic-flux maps. 

As demonstrated by \citet{gon07}, the Sun's far-side acoustic phase shifts, the physical quantity displayed in helioseismic far-side images, are also well correlated with magnetic-field strengths. 
This can also be seen from Figure~\ref{composite_img}b and \ref{composite_img}d, where the locations, sizes, and amplitudes of strong acoustic phase deficits, exhibited as dark patches in the figure, seem to show a good correlation with those of the far-side 304\,\AA\ brightening areas. 
We therefore can train a second machine-learning model by coupling the far-side magnetic-flux maps, which are converted from the \euvi\ 304\,\AA\ images using the first trained machine-learning model, and the far-side helioseismic images. 
The resulting model is able to convert the far-side helioseismic images into far-side magnetic-flux maps. 

Eventually, this second machine-learning model, together with our time--distance helioseismic far-side imaging model that was previously developed \citep{zha19}, is able to generate far-side magnetic-flux maps by { only using}  continuous near-side full-disk Doppler-velocity observations.
Such far-side magnetic-flux maps can be routinely processed in \hmi's data-processing pipeline.
This is a substantial progress that would help the space-weather forecasting and coronal-field modeling, and lower our reliance on off-Sun-Earth-line observations for far-side magnetic fields.

\begin{figure}[!ht]
\centering
\includegraphics[width=0.9\textwidth]{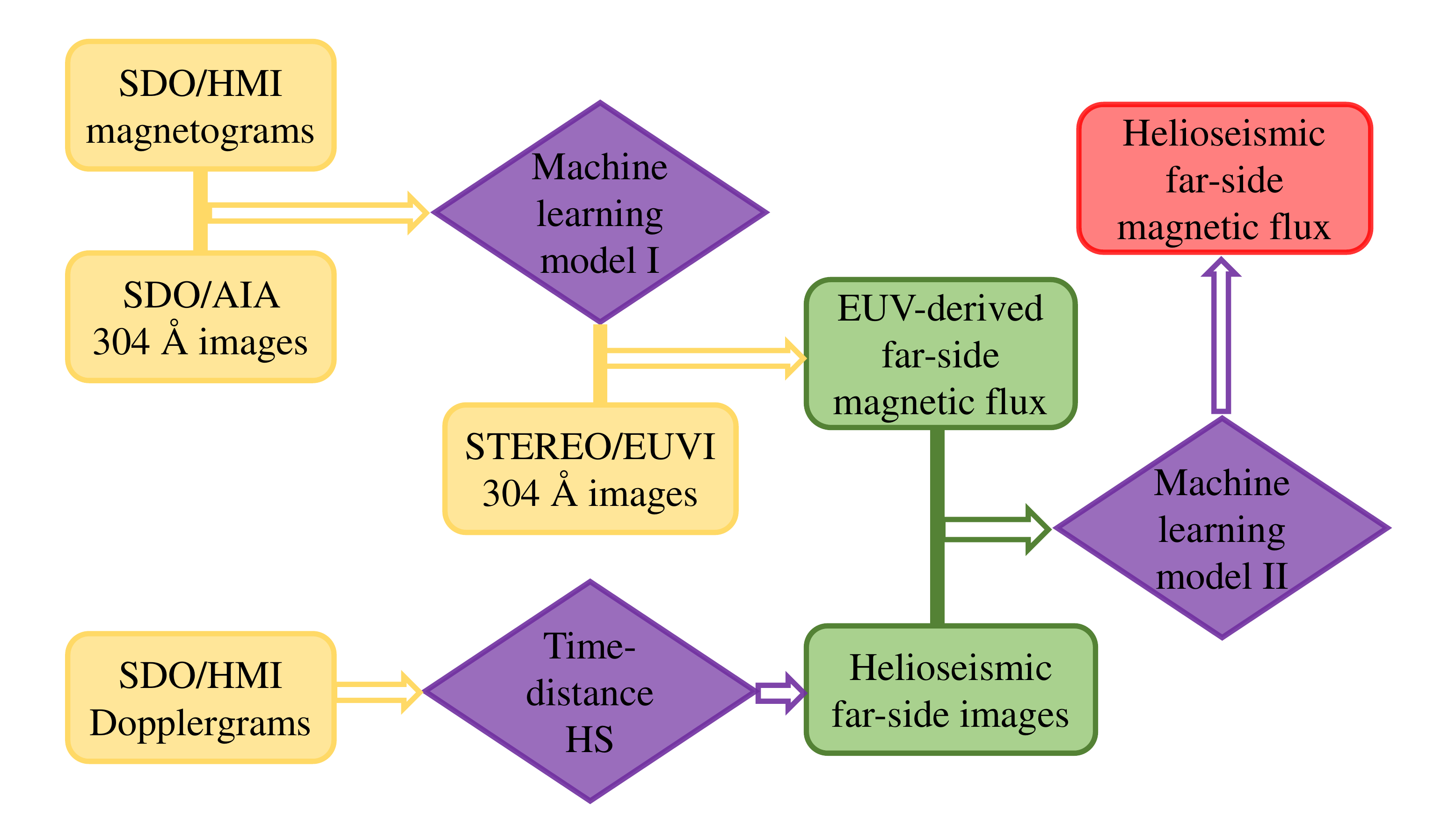}
\caption{Flow chart showing the entire procedure of making far-side magnetic-flux maps using the near-side Doppler-velocity observations.
Yellow boxes represent original data inputs, green boxes represent intermediate data products, and the red box represents our final data product. 
Purple diamonds represent the methods that are used in the corresponding processes.}
\label{flow_chart}
\end{figure}

Summarizing the analysis procedure prescribed above, Figure~\ref{flow_chart} presents the flow chart of our data preparation, codes { development}, and machine-learning training.
The original data inputs, shown as yellow boxes in this flow chart, include \hmi\ near-side magnetograms and simultaneous \aia\ near-side 304\,\AA\ images, both of which are used in the first machine-learning model training, the \hmi\ near-side Dopplergrams that are used by a time--distance helioseismic code for the Sun's far-side helioseismic images, and the \euvi\ 304\,\AA\ images that are used to bridge the near and far sides. 
The intermediate products, shown as green boxes, include EUV-derived far-side magnetic-flux maps and far-side helioseismic images.
Our final data product from this work, shown as the red box, is the { maps of the far-side unsigned magnetic flux} derived from the helioseismic images.

\subsection{Data Preparation: \hmi\ and \aia\ Data}
\label{sec22}

To prepare the data that are used in our first machine-learning training, for each day from 2010 May 1 through 2019 April 30, we take 4 pairs of simultaneously-observed (with less than 24\,s time difference) \hmi\ line-of-sight magnetograms and \aia\ 304\,\AA\ images. 
A total of $\sim$13,000 images are selected during this 9-year period. 
Corrupted images and data captured during flares are excluded.
In each calendar year, the data in January are not used in the machine-learning training but are kept for testing the trained model.
More specifically, the data during January 1 and 16 { of each year}, about 4\% of the total dataset are reserved as the test set for final evaluation, and the data during January 17 and 31 are reserved as the development set for the performance evaluation in the training process. 
The data between February 16 and December 16, about 83\% of the total dataset, are used in the machine-learning training. 
Fifteen days of data before the start and after the end of each training data period, i.e., February 1 -- 15 and December 17 -- 31, are unused in either training or testing to prevent the same ARs from appearing in both the training and testing sets. 

The line-of-sight magnetograms are averaged from three consecutive \hmi\ 720\,s magnetograms, and the 304\,\AA\ images are averaged from five consecutive images with a 24\,s cadence.
For each pair of images, the off-disk data in the 304\,\AA\ image are truncated, and both the magnetogram and 304\,\AA\ image are rebinned to 512$\times$512 pixels and spatially aligned, with the alignment accuracy no worse than 1.0$\arcsec$. 

\begin{figure}[!ht]
\centering
\includegraphics[width=0.99\textwidth]{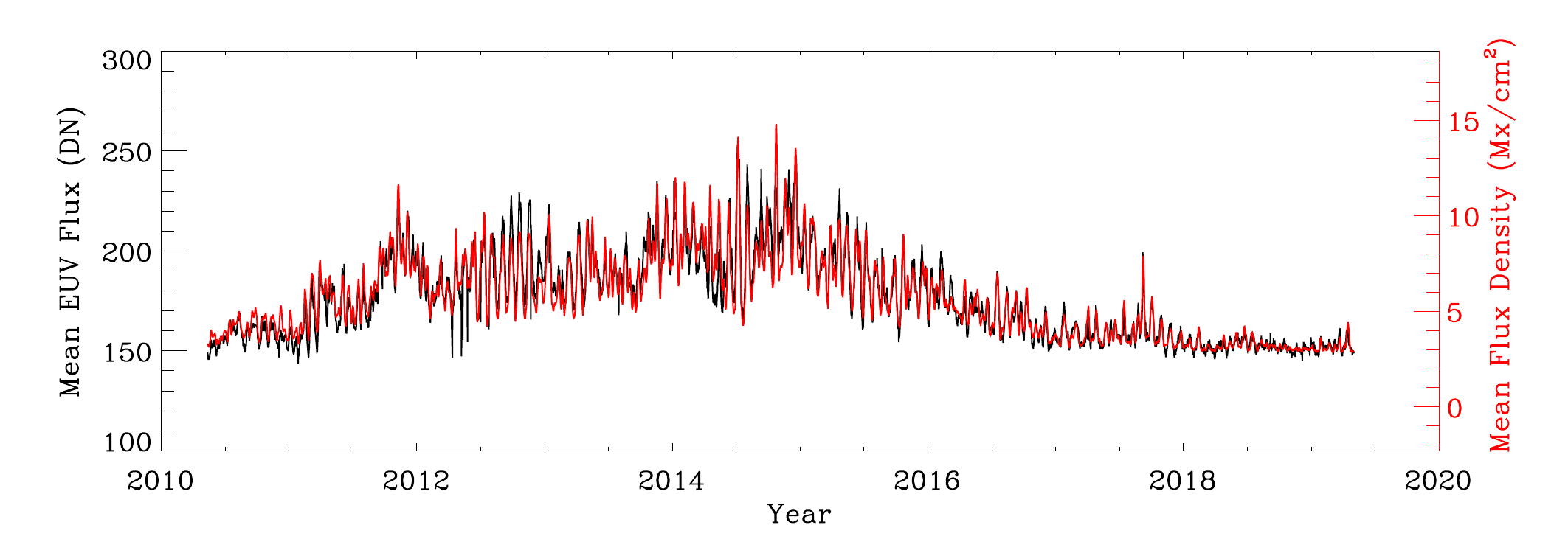}
\caption{Comparison of mean \aia\ 304\,\AA\ flux and mean \hmi\ line-of-sight magnetic flux density, after degradation of the \aia\ 304\,\AA\ data is recalibrated.}
\label{AIA_los_flux}
\end{figure}

The \aia\ 304\,\AA\ data need to be recalibrated before being used for the machine-learning training.
As noted by many authors \citep[e.g.,][]{boe14, gal19}, due to the instrumental degradation and the exposure time changes, the flux density in the 304\,\AA\ images varied substantially with time through the selected 9-year period. 
The \aia\ team proposed an empirical recalibration method\footnote{\url{https://www.lmsal.com/sdodocs/doc/dcur/SDOD0060.zip/zip/entry/sdoguidese7.html\#x12-600007.6}} so that the flux density remains relatively stable.
However, the solar spectral irradiance measured by the \sdo/EVE \citep[Extreme Ultraviolet Variability Experiment;][]{woo12} MEGS-A detector, which was used for the \aia\ EUV recalibration, became unavailable after mid-2014 because of a power anomaly. 
While the \aia\ team afterwards used an alternative scheme involving the irradiance provided by the Flare Irradiance Spectral Model \citep{cha07, cha08}, we choose to recalibrate the \aia\ 304\,\AA\ intensity by using the \hmi\ mean magnetic-flux density. 
The EUV flux level calibrated before the loss of \sdo/EVE MEGS-A follows a good linear relation with the \hmi\ mean flux, we therefore apply a piece-wise empirical correction on the EUV 304\,\AA\ flux based on the magnetic-flux density after the MEGS-A loss, so that the corrected 304\,\AA\ flux follows a same linear relation with the magnetic flux throughout the interested period.
Figure~\ref{AIA_los_flux} shows the corrected 304\,\AA\ mean flux in comparison with the \hmi's mean magnetic-flux density.

\subsection{Data Preparation: \stereo/EUVI data}
\label{sec23}

Both \stereo-A and \stereo-B spacecraft gradually shifted positions relative to the earth as they continued their journey around the Sun. 
During the period when \stereo-A, \stereo-B, and \aia\ observations are all available --- between 2010 May 13 and 2014 August 16, the combined observations from both \stereo\ spacecraft cover the entire far side of the Sun for about 4.3 years. 
In case of partial coverage that only happened in the first few months, the missing portions are left blank and not used in the training.
Many authors combined these data into instantaneous synchronic charts in their analysis, \citep[e.g.,][]{cap16, lie17, zha19}; however, in this work, in order to keep these \euvi\ images appearing similar to the \aia\ images, i.e., in apparent disk images, so that our first machine-learning trained model can be applied directly on these observations, we keep the \stereo-A and \stereo-B 304\,\AA\ images separate without any coordinate transformation or image combination.

\begin{figure}[!ht]
\centering
\includegraphics[width=1.0\textwidth]{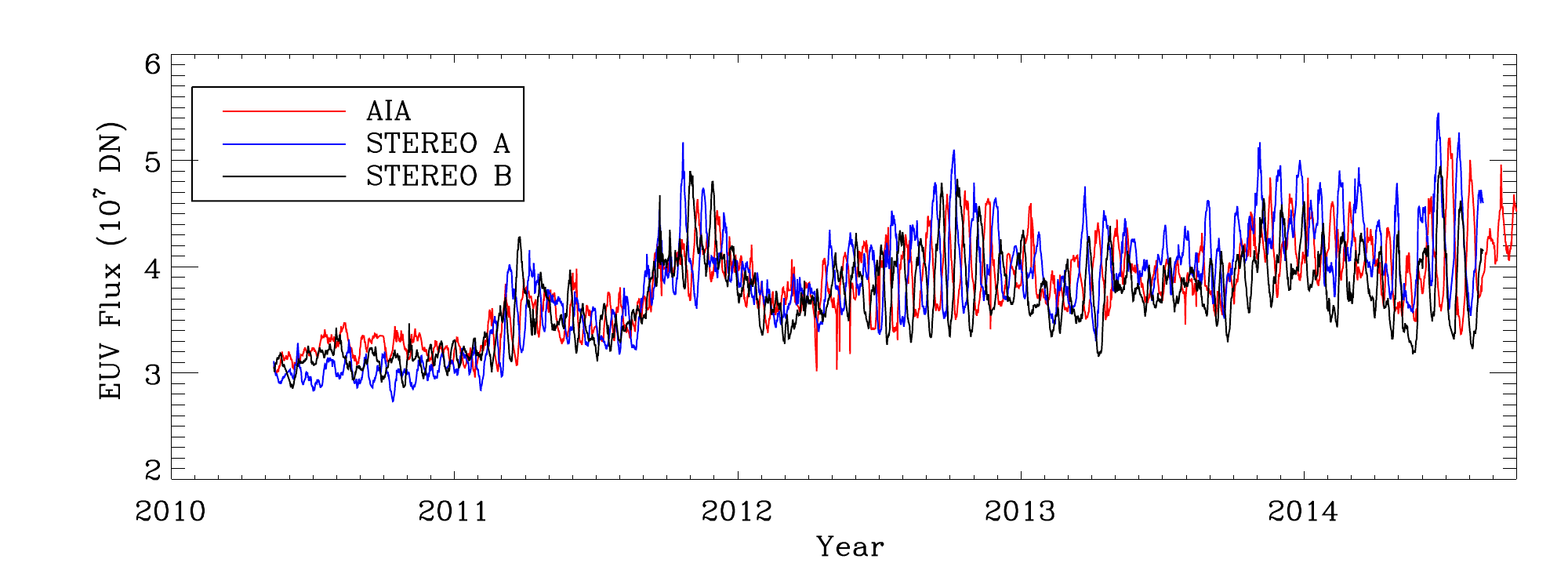}
\caption{Comparison of EUV 304\,\AA\ total fluxes observed by the \aia\ and the two \euvi\ instruments during the period of interest, after the \euvi\ images are corrected for the instrumental degradation and recalibrated to match \aia\ values.}
\label{AIA_EUVI_flux}
\end{figure}

However, similar to the \aia, both of the \stereo-A and \stereo-B EUVI instruments also suffered gradual degradation, and the two instruments might have experienced different levels of degradation. 
Therefore, before these \euvi\ 304\,\AA\ images are converted into magnetic-flux maps, they also need to be corrected for the degradation; and, to keep the data consistent with the \aia\ observations, these data also need to be recalibrated to match the \aia\ 304\,\AA\ flux.
In order to carry out such a correction and recalibration, we first make 27-day running averages on the \stereo-A and \stereo-B and \aia\ total fluxes. 
Assuming that the temporal variation of the averaged total fluxes is slow and that the total fluxes observed by the three instruments are consistent with each other with just a few days of time delay due to different vantage points, we correct the degradation of the two \euvi\ instruments and recalibrate their 304\,\AA\ values matching those of the \aia\ with a time-dependent piece-wise scaling factor. 

Figure~\ref{AIA_EUVI_flux} shows a comparison of our recalibrated \euvi\ total EUV fluxes with the \aia\ total EUV flux without the 27-day averaging.
These recalibrated \euvi\ images are then used in our second machine-learning training in Section~\ref{sec4}.

\subsection{Data Preparation: Helioseismic Far-Side Images}
\label{sec24}

The far-side helioseismic data used in our second machine-learning training are time--distance helioseismic results that are routinely generated by an \hmi's pipeline \citep{zha19}. 
One far-side helioseismic image is calculated every 12 hours using 31-hr-long 45-sec cadence \hmi\ full-disk Doppler-velocity data.
Every far-side helioseismic image used in this work is an average of seven far-side 12-hr-cadence images, including the current one and six immediate prior ones, for a higher signal-to-noise ratio.
Due to the rotation of the Sun, the far-side west-limb areas are less averaged than the other areas, hence being noisier.
These far-side images are calculated in longitude -- sin(latitude) coordinates, with a spatial sampling rate of 0$\fdg8$ pixel$^{-1}$ { in longitude and 0.0144 pixel$^{-1}$ for sin(latitude)}, but the actual spatial resolution of such images are believed to be close to the wavelength of the helioseismic waves, which is around 80\,megameters.
The time--distance far-side images cover the period of 2010 May 1 to the present, a total of 12 years with over 8,700 images available; however, due to the limitation of \stereo's far-side coverage, only 4.3 years of the data will be used in the machine-learning training and testing.

\section{Machine-Learning Model I: Converting EUV Images into Magnetic-Flux Maps}
\label{sec3}

\subsection{Machine-Learning Training I}
\label{sec31}

\begin{figure}[!ht]
\centering
\includegraphics[width=0.8\textwidth]{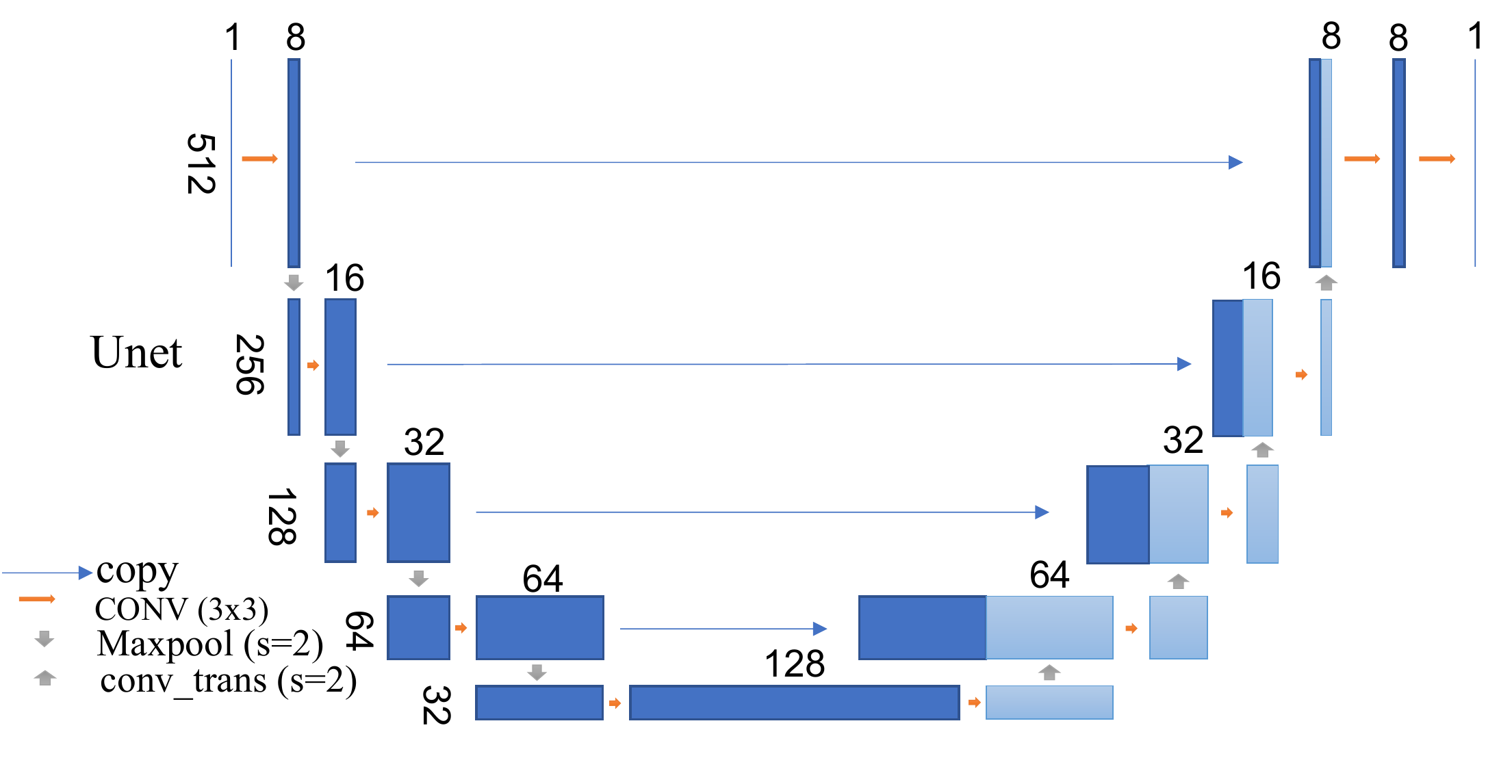}
\caption{Unet deep neural network architecture for the model converting EUV 304\,\AA\ images into magnetic-flux maps.}
\label{unet_training1}
\end{figure}

Our first machine-learning training pairs the near-side observations of magnetic-flux (unsigned magnetic-field) maps and EUV 304\,\AA\ images that are prepared in Section~\ref{sec22}.
A deep convolutional neural network (CNN) model \citep[e.g.,][]{fuk82, lec89} { is} trained to map a magnetic-flux map to an EUV map, by convolving the input magnetic-flux map with a series of convolutional kernels, each of which is followed by a nonlinear mapping. 
The combination of a convolutional transformation and one nonlinear mapping is called one convolutional layer. 
The machine-learning training process is an optimization problem to search for optimal kernel values to minimize a loss function, which is defined to quantify the difference between produced maps and target maps. 
Here we use a CNN with Unet architecture \citep[e.g.,][]{ron15}, with its channel number, kernel size, and the image size of each layer as illustrated in Figure~\ref{unet_training1}. 
A rectified linear unit (ReLU) nonlinear mapping is applied after each 3$\times$3 convolution, followed by a 2$\times$2 maxpool down-sampling. 
After each down-sampling step, feature channels are doubled. 
The image is eventually downsized to $32\times32$ pixels with 128 channels, after which convolutional up-sampling is applied. 
Between each up-sampling step, we concatenate the up-sampled and previously down-sampled channels of the convolutional layer with the same image size, and perform a 3$\times$3 convolution followed by a ReLU nonlinear mapping. 
After each up-sampling step, feature channels are halved. 
The final output map has same dimensions as the original input map. 

Both the \hmi\ magnetic-flux maps and \aia\ 304\,\AA\ images are originally tracked and processed in the form of disk images. 
However, in the machine-learning training, they are remapped into longitude -- sin(latitude) coordinates, so that the data fill all of the 512$\times$512 grids and margins beyond the solar limb can be removed, which otherwise increase the computational burden without useful scientific returns. 
This also takes full advantage of the convolutional neural network, avoids artifacts in the convolution across limbs, and preserves better the shapes of ARs when they are close to either of the limbs. 
Moreover, because the Sun is a sphere, any type of remapping will cause the inhomogeneity of pixel sizes and geometry across the remapped images. 
Longitude -- sin(latitude) coordinates are a good choice since these maps are uniform across longitudes and change relatively slowly across latitudes, particularly in the range of active latitudes where this work is more concerned.

We apply gradient descent with the Adam algorithm to minimize the loss function \citep{Adam}, which is defined as the pixel-to-pixel mean squared error between the prediction images and the target images for the whole training set.
The high-flux regions on the images are given a larger weight, since they are the extreme cases of the training data distribution and are otherwise less visited in the training.
The metrics used for evaluation and selection of the trained models are the mean squared errors and the linear correlation coefficients between the prediction images and the target images for the development sets, which are reserved during the data preparation (Section~\ref{sec22}).

\subsection{Testing of the Trained Model I}
\label{sec32}

\begin{figure}[!ht]
\centering
\includegraphics[width=1.0\textwidth]{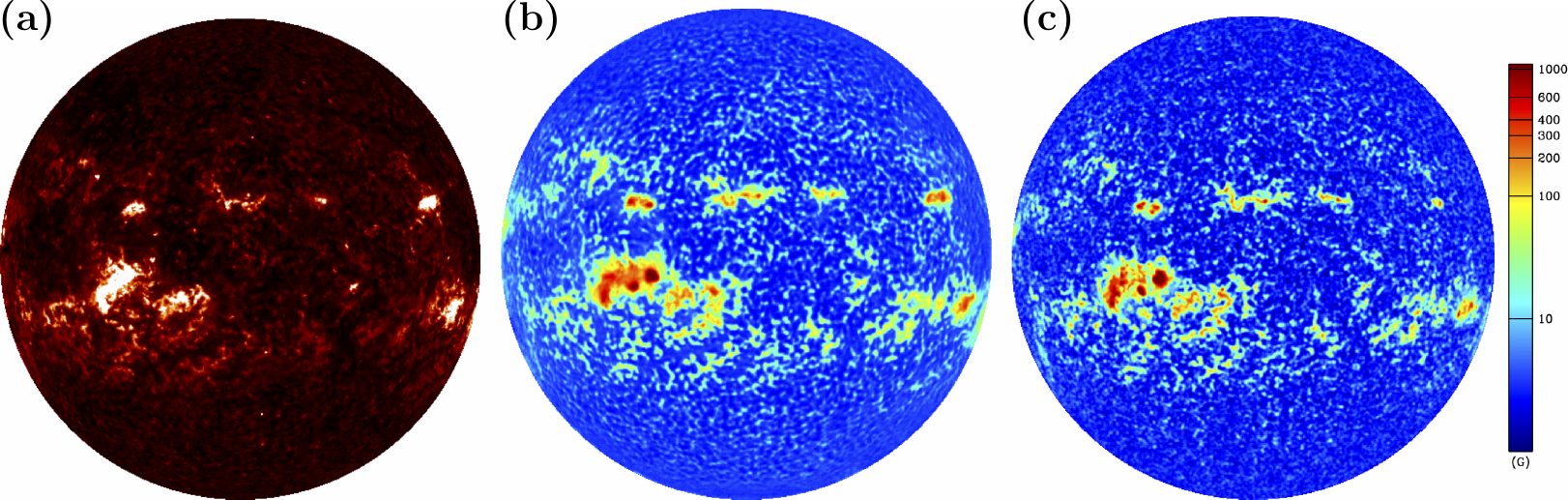}
\caption{(a) An \aia\ 304\,\AA\ image of the Sun, taken at 00:00\,UT of 2014 January 6; (b) AI-generated magnetic-flux map based on the 304\,\AA\ image shown in panel (a); (c) Simultaneous \hmi-observed magnetic-flux map after a $3\times3$ boxcar smoothing for a comparison with panel (b).}
\label{training1}
\end{figure}

After the first machine-learning model is trained, we apply the trained model on the test datasets that are put aside during the data preparation. 
Figure~\ref{training1} shows a selected comparison between the AI-generated magnetic-flux map, converted from an \aia\ 304\,\AA\ image, and the simultaneous \hmi-observed unsigned magnetic-flux image.
It can be seen that, globally, the AI-generated magnetic-flux map has reproduced all magnetic structures that were observed by the \hmi, with a one-to-one correspondence; and locally, each active region has similar magnetic structures compared with the observed magnetic-flux map.

\begin{figure}[!ht]
\centering
\includegraphics[width=0.45\textwidth]{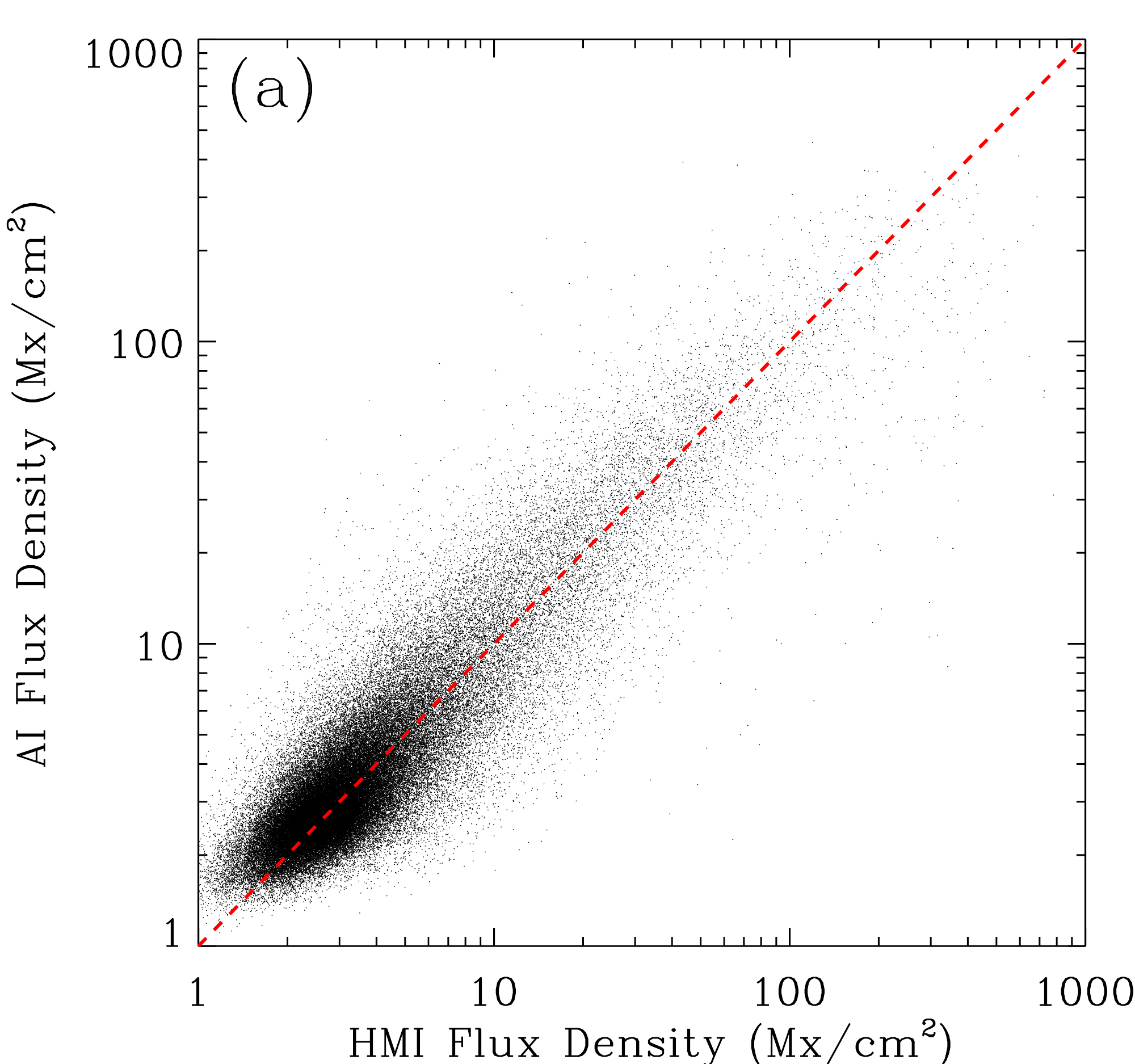}
\includegraphics[width=0.45\textwidth]{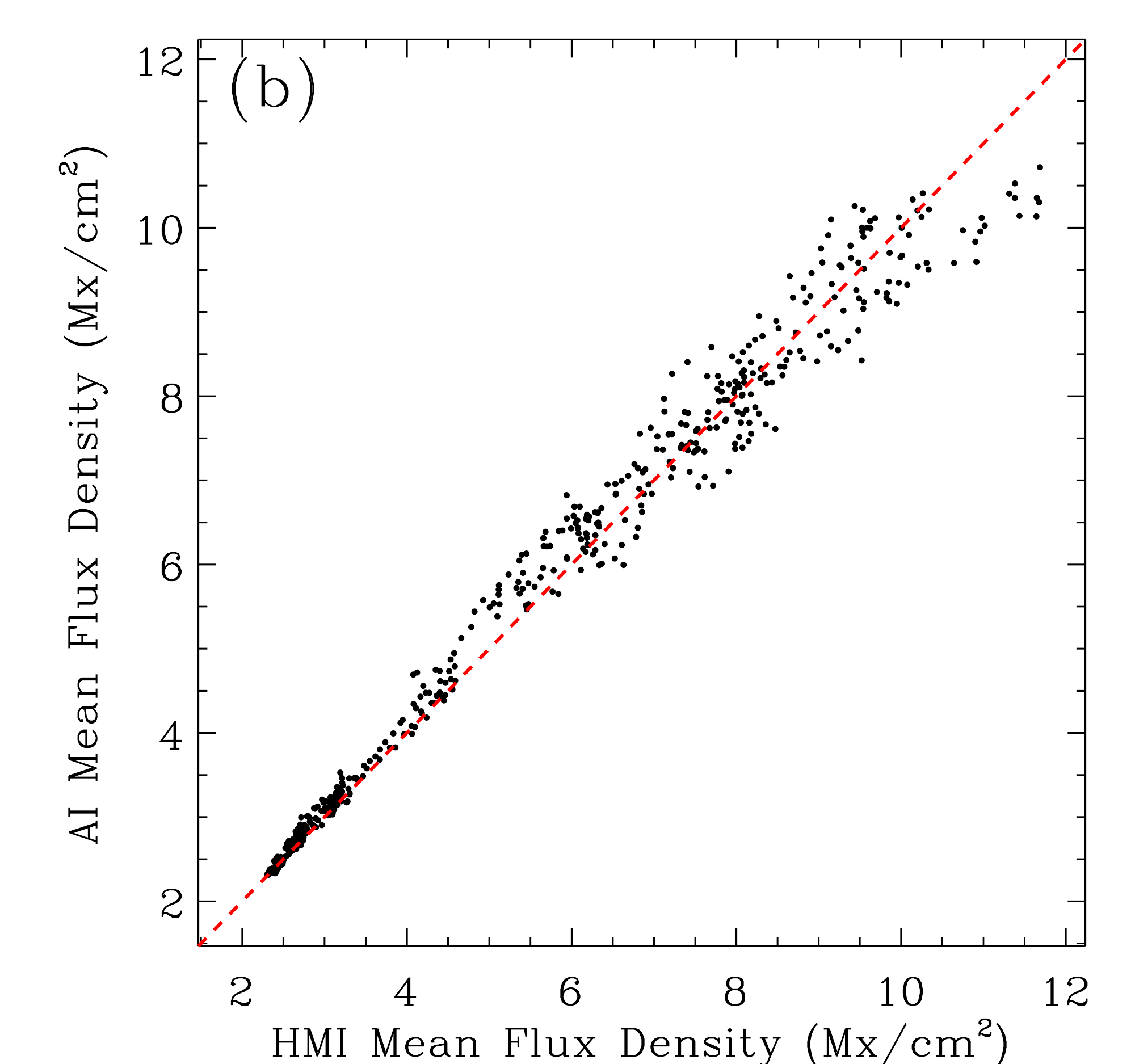}

\caption{(a) Scatter plot comparing the values of magnetic-flux densities from the AI-generated maps and those from the \hmi-observed maps. 
(b) Scatter plot comparing the globally-averaged mean magnetic-flux densities from the AI-generated maps and those computed from the \hmi-observed images. 
One point in the plot represents the results from one image in the test dataset.}
\label{scatter1}
\end{figure}

While Figure~\ref{training1} shows the remarkable morphological resemblance between the observed magnetic-flux maps and AI-generated maps, Figure~\ref{scatter1} further shows how the values of magnetic-flux densities compare.
The scatter plot in Figure~\ref{scatter1}a shows a pixel-to-pixel comparison of all the test datasets, covering the nine 15-day periods with 4 pairs of datasets each day, after an 8$\times$8-rebin of each image. 
The correlation coefficient between these rebinned images is as high as 0.88, and the ratio between the two sets of values is close to 1.0.
The scatter plot in Figure~\ref{scatter1}b shows the comparison of the globally-averaged mean magnetic-flux densities, obtained for each pair of AI-generated images and directly-observed images.
A correlation coefficient of 0.99 is found between these two sets of averaged fluxes.

As can be seen from these comparisons, while keeping the general shapes and structures of the EUV 304\,\AA\ images, the AI-generated magnetic-flux images show very high correlation with the observed magnetic-flux images in both spatial structures and the values of magnetic flux. 
Despite the high correlation between the EUV-based AI-generated magnetic-flux maps and the \hmi-observed maps, one also needs to recognize that in our trained flux maps, the high flux values often tend to be lower than the observed ones. 
This is probably due to the convolution calculations used in the machine learning, which smear out high values and smooth the data. 
This is also likely due to the EUV flux being saturated for strong active regions.
Meanwhile, we also need to emphasize that the relation between the EUV emission intensity and magnetic-flux density is only an approximate power-law relation rather than being a strict one \citep{uga15}. 
Other than that, the AI-generated maps are very good proxies to magnetic-flux maps when the latter are not directly observable.

\subsection{Applying Machine-Trained Model on \euvi\ Data}
\label{sec33}

\begin{figure}[!b]
\centering
\includegraphics[width=1.0\textwidth]{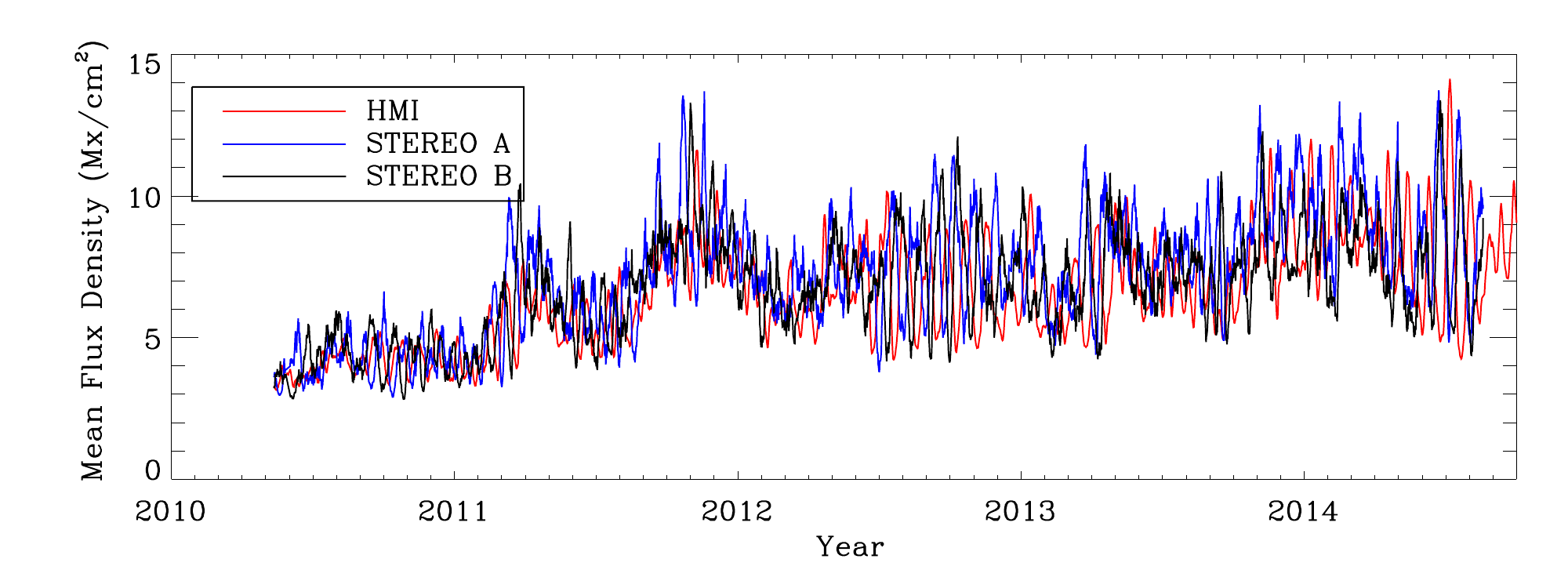}
\caption{Comparing the globally-averaged mean magnetic-flux densities, computed from the \hmi-observed magnetic-flux maps and the magnetic-flux maps converted from two sets of \stereo-observed 304\,\AA\ images.}
\label{comp_mag}
\end{figure}

The machine-trained model is then applied on the \euvi-observed 304\,\AA\ images, which are prepared separately for the observations by \stereo-A and \stereo-B following prescriptions in Section~\ref{sec23}. 
Figure~\ref{comp_mag} shows a comparison of globally-averaged mean magnetic-flux densities, computed from the \hmi-observed magnetic-flux maps and the magnetic-flux maps converted from the \stereo-A and \stereo-B images, respectively, by the machine-trained model with a 12-hr cadence covering the period of May 2010 to October 2014. These curves are not expected to be identical because they observe different parts of the Sun, but their general trends are very similar and their amplitudes are also close, demonstrating the reasonableness of our magnetic-flux estimates from the \stereo-observed 304\,\AA\ images.

\begin{figure}[!t]
\centering
\includegraphics[width=1.0\textwidth]{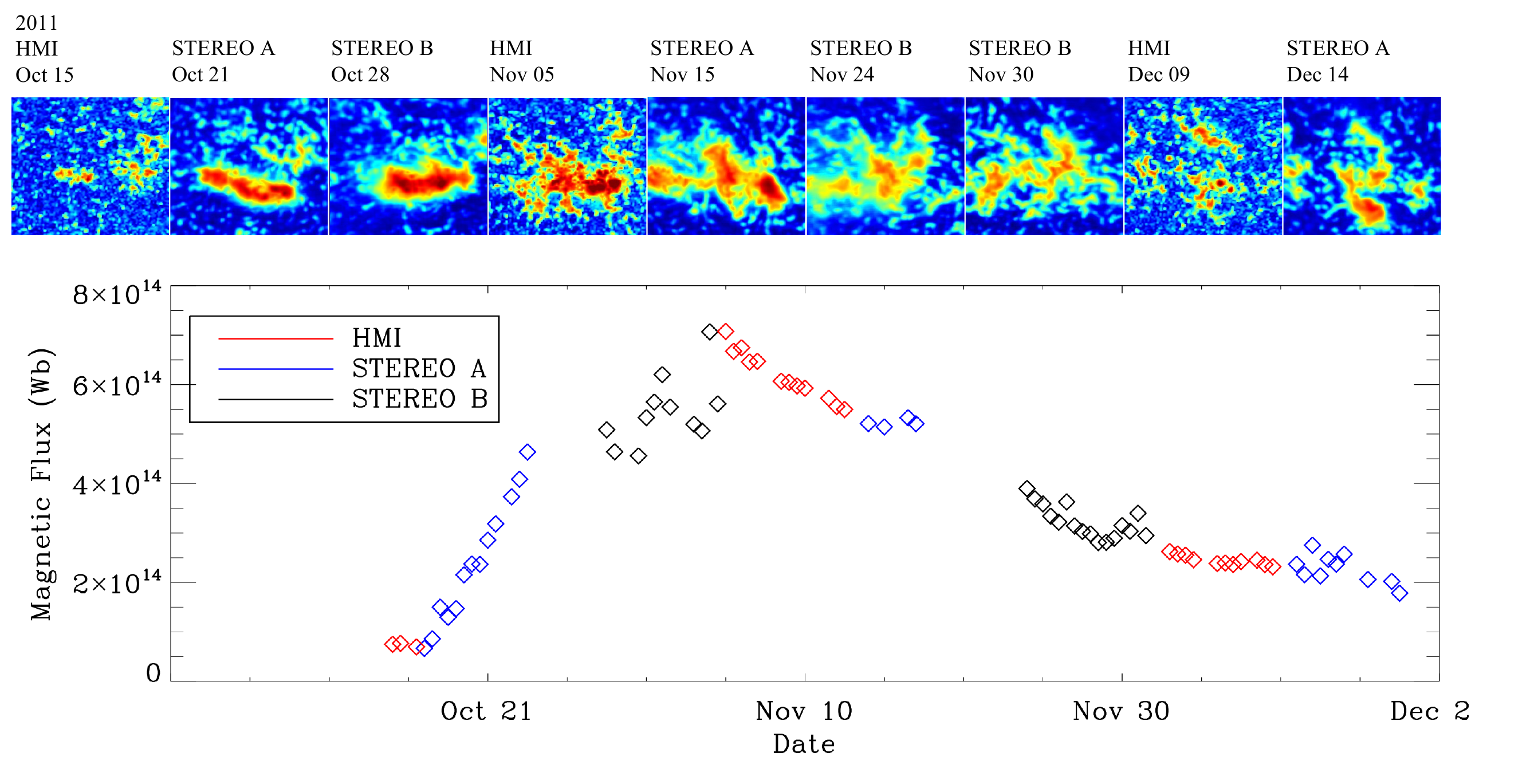}
\caption{Temporal evolution of the total magnetic flux for NOAA AR\,11339 during its multiple disk passages on both near and far sides of the Sun.
}
\label{AR_track}
\end{figure}

Figure~\ref{AR_track} shows another examination of the AI-generated far-side magnetic-flux maps --- the full life of a large active region. 
The NOAA AR\,11339 emerged on the Sun's near side around 2011 October 15, and finally decayed away two months later around 2011 December 14 on the Sun's far side. 
It had three times of near-side disk passage in the \hmi's field of view, three passages in the \stereo-A's field of view, and two passages in the \stereo-B's field of view. 
A combination of direct observations and AI-converted magnetic-flux maps show the temporal evolution of this region nearly uninterruptedly (with some exceptions when the AR was near the limbs, i.e., $\ge 60\degr$ from the respective disk center, of both \hmi\ and \euvi) for its entire lifetime, and the total flux calculated for this region shows its rapid growth and slow decay, associated with sporadic small-scale flux emergence during the period. 
Such coverage of magnetic-flux maps for a large AR over its entire life span, as the one shown in Figure~\ref{AR_track}, has never been achieved before. 

\section{Machine-Learning Model II: Converting Helioseismic Far-side Images into Magnetic-Flux Maps}
\label{sec4}

\subsection{Preparation of Far-side Magnetic-Flux Maps}
\label{sec41}

Our second machine-learning model { is trained to} convert helioseismic far-side images, which are measurements of acoustic waves' phase deficits { arising in the vicinity of} far-side ARs, into far-side magnetic-flux maps through pairing with the \stereo-converted magnetic-flux images obtained in Section~\ref{sec33}. 
However, before carrying out this machine-learning training, we need to further process the \stereo-converted magnetic-flux maps. 
The 304\,\AA\ images observed by both \aia\ and the two \euvi\ instruments are all affected by the limb-darkening effect as well as disk projections, the former causing the EUV emission intensity gradually lower toward the limbs and the latter affecting the total visible areas and the line-of-sight component of magnetic fields. 
Because the near-side machine-learning training is done between 304\,\AA\ images and line-of-sight magnetic-flux maps without these effect corrections, the \stereo-converted magnetic-flux maps carry these limb-darkening and line-of-sight projection effects in the same way as the observed line-of-sight magnetic-flux maps.
When combining magnetic-flux maps generated from the two \stereo\ instruments into one far-side magnetic-flux map, the observed disk limbs are often not or even near the limbs of the Sun's far side; therefore corrections are needed to compensate the effects from the limb darkening and the projection.  
An empirical method, through weighted averaging of seven synthetic far-side magnetic-flux maps as if observed from seven different vantage points, has been developed to make these corrections. 
However, the detailed process of this correction is beyond the scope of this article, thus is omitted here.

Once the limb-darkening and disk-projection effects are empirically corrected in the \stereo-generated magnetic-flux maps, we remap these disk images from the two spacecraft into Carrington coordinates, combine them into one bigger map, and then further crop and transform them into longitude -- sin(latitude) coordinates to match the helioseismic far-side images with identical scales and grids of 225$\times$121.
This coordinate system uses the Sun's apparent disk coordinate instead of the heliographic coordinate, and this way, we do not need to consider effects and coordinate transformations due to the $B$-angle variations.
Twice every day during the 4.3-year period, two \stereo-converted magnetic-flux maps are thus combined into one full far-side image in the new coordinate system. 
Note that with the continuing movement of both \stereo\ spacecraft, the empirical correction of the flux maps and the merger of the two images need to be adjusted accordingly.

\begin{figure}[!ht]
\centering
\includegraphics[width=1.0\textwidth]{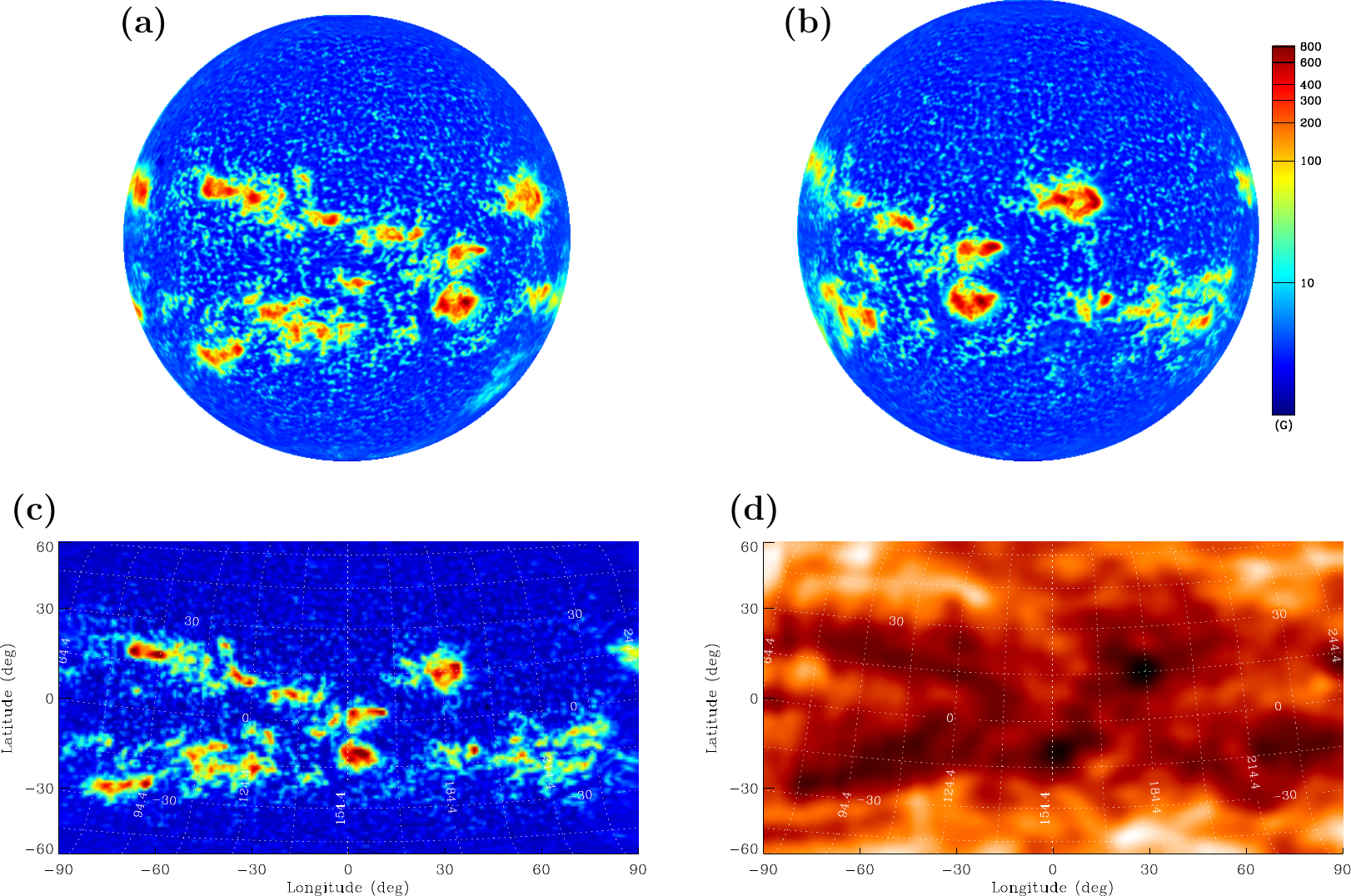}
\caption{(a) Magnetic-flux map converted from \stereo-A 304\,\AA\ image at 00:00UT of 2014 March 13. 
(b) Same as panel (a) but for \stereo-B converted magnetic-flux map. 
(c) Far-side magnetic-flux map merged from both maps in panels (a) and (b).
(d) Helioseismic far-side image taken at the same time as the map shown in panel (c).
Both maps shown in panels (c) and (d) are in the same longitude -- sin(latitude) coordinate system, where the longitude and latitude are the apparent disk coordinate. 
The heliographic coordinates are marked as dashed lines in both panels. }
\label{data_ready}
\end{figure}

Figure~\ref{data_ready} shows an example of magnetic-flux maps converted from observations of both \euvi\ instruments, the merged far-side flux map from these two separate maps, and the simultaneous helioseismic far-side image.
These merged far-side magnetic-flux maps and the helioseismic far-side images in the same coordinate system, as displayed in Figure~\ref{data_ready}c and \ref{data_ready}d, are then used for our second machine-learning training.

\subsection{Machine Learning Training II}
\label{sec42}

The far-side magnetic-flux maps, merged together using the AI-generated magnetic-flux maps from both \euvi\ observations, are paired together with the far-side helioseismic images for our second machine-learning training for a model that converts far-side helioseismic images into { maps of far-side unsigned magnetic flux}.
Two pairs of data are available each day covering the period of 4.3 years, totaling about 3,100 pairs of data for our training and testing.
The data in March of each year, about 8\% of the total, are kept for the final testing of the model. 
There are only 240 pairs of images available for testing, so we do not further split the data into the test set and development set as we do for the first machine-learning training. 
Still, 15 days of data immediately before and after the testing period are discarded to avoid duplicate ARs in the training and test sets; and the rest of the data, about 83\% of the total, are used for the machine-learning training.

\begin{figure}[!ht]
\centering
\includegraphics[width=0.8\textwidth]{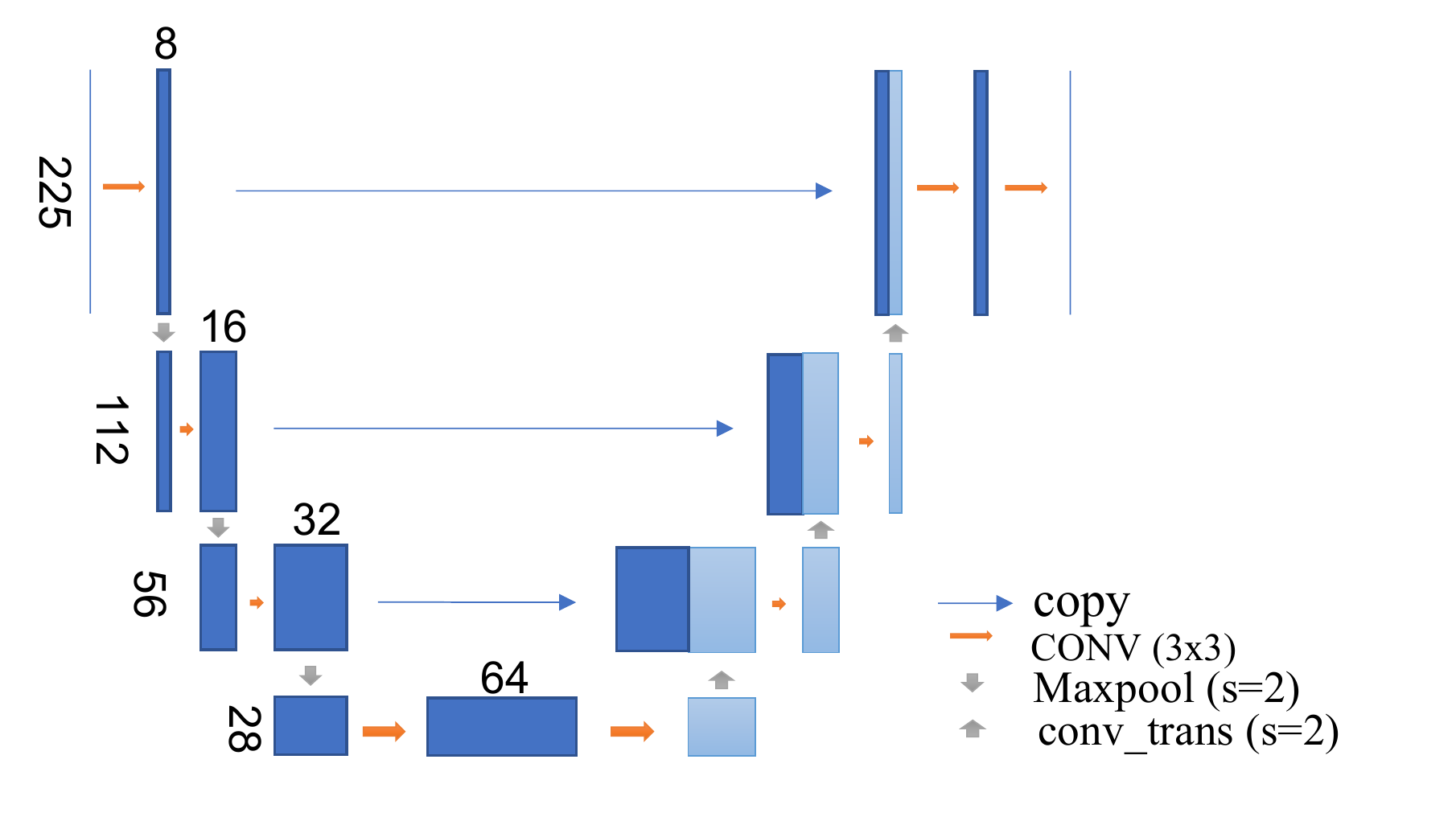}
\caption{Unet deep neural network architecture for the second machine-learning training, converting far-side helioseismic images into magnetic-flux maps.}
\label{unet_training2}
\end{figure}

In this machine-learning training, we use a Unet neural network architecture similar to the one used in our first training (Section~\ref{sec31}).
However, it should be recognized that although the far-side helioseismic images have a temporal cadence of 12 hours, these data are actually averaged from 3.5 days of observations, making the effective number of independent datasets even smaller.
Due to the relatively small training set and the lower dimensionality of the images, in this training we use fewer convolutional layers in the framework than those used in the first training.
The architecture of the neural network is shown in Figure~\ref{unet_training2}.
The loss function, metrics, and optimization algorithm are chosen to be the same as those used in the first neural network training (Section~\ref{sec31}). 

\subsection{Testing of the Trained Model II}
\label{sec43}

\begin{figure}[!ht]
\centering
\includegraphics[width=0.93\textwidth]{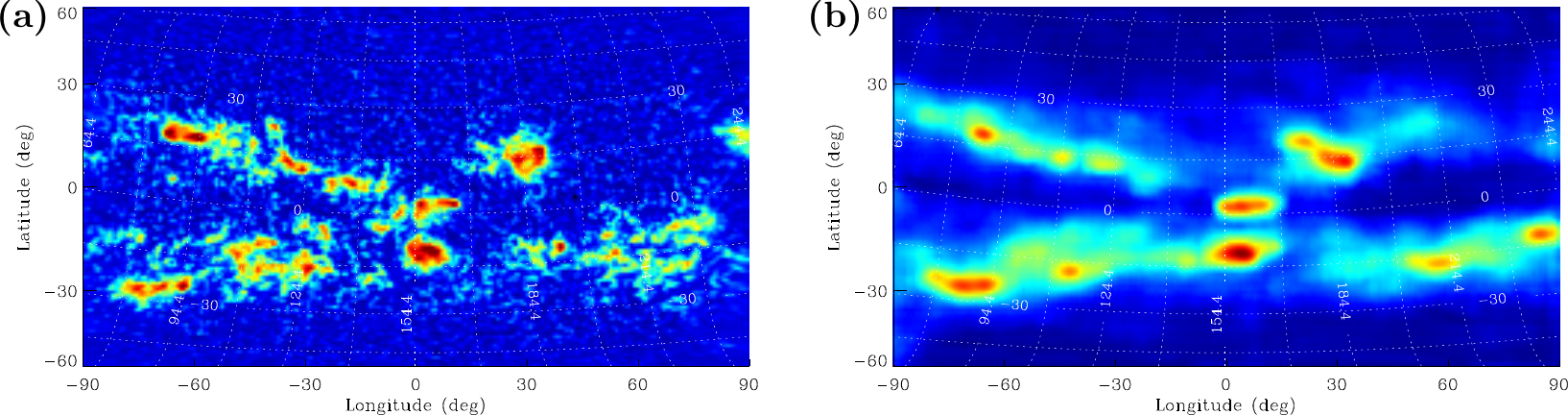}
\includegraphics[width=0.056\textwidth]{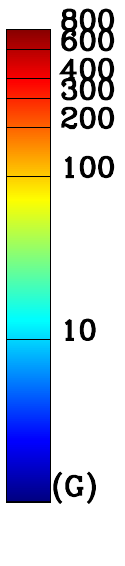}
\caption{Example showing the comparison of (a) \stereo-converted far-side magnetic-flux map and (b) helioseismic-image-converted far-side magnetic-flux map, both taken at 00:00\,UT of 2014 March 13.}
\label{training2}
\end{figure}

After our second machine-learning model is trained, it is applied on the reserved test dataset for a performance evaluation.
Figure~\ref{training2} shows a comparison of a selected \stereo-converted magnetic-flux map, which serves as ground truth in this case (although they are actually also AI products), and the corresponding helioseismic-image-converted magnetic-flux map. 
It is worth being reminded that because only low harmonic-degree data are used in the helioseismic-imaging technique, the far-side images obtained from this method have low spatial resolution; and the high noise in the helioseismic-imaging method often cause some spurious signals in the far-side images. 
Therefore, the resultant helioseismic-image-converted magnetic-flux maps have low spatial resolution, and are often associated with spurious regions that may not correspond to a true far-side AR. 
Nevertheless, the comparison in Figure~\ref{training2} shows that the helioseismic-converted flux maps are able to capture most of the large ARs with a reasonable accuracy in the magnetic-flux density values. { The correlation between the two panels in Figure~\ref{training2} is 0.78 after the spatial resolution of the \stereo-derived map is lowered to match that of the helioseismic image.  The mean correlation between the \stereo-derived magnetic-flux maps and the helioseismology-derived maps from the entire test data set is $0.41\pm0.16$, and this correlation coefficient varies significantly with the solar cycle, larger with higher solar activity. For active days when the mean flux density is larger than 8 Mx  cm$^{-2}$, the mean correlation is $0.55\pm0.13$.} 

\begin{figure}[!ht]
\centering
\includegraphics[width=0.6\textwidth]{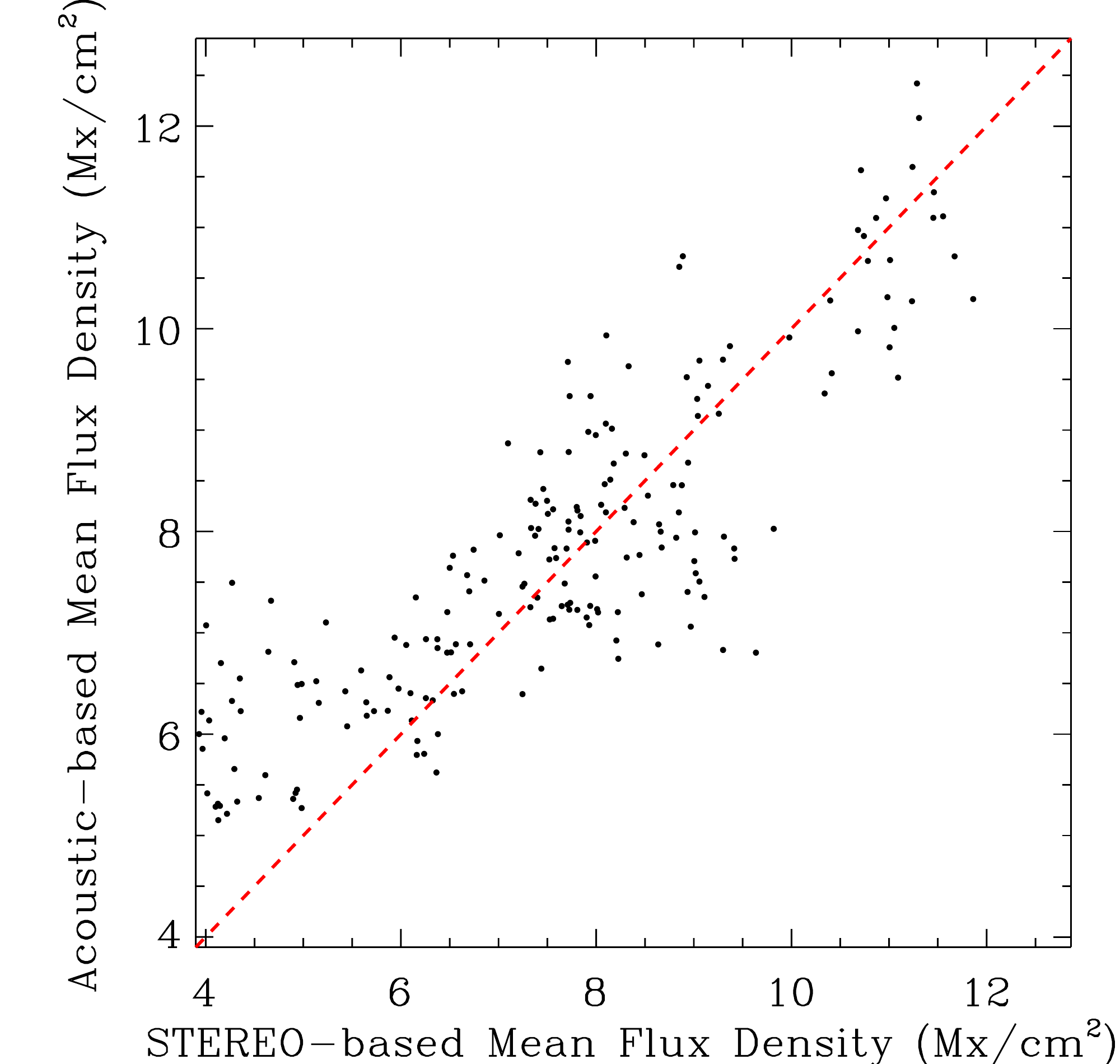}
\caption{Scatter plot comparing the mean magnetic-flux densities from the \stereo-generated maps with those from the helioseismic-image-converted maps, both of which are from the test dataset.}
\label{scatter2}
\end{figure}

Figure~\ref{scatter2} shows the scatter plot comparing the \stereo-converted and the helioseismic-image-converted magnetic fluxes, with each point in the plot averaged from the entire covered far-side area in each test dataset. 
The mean flux-density values are relatively wide spread, with a correlation coefficient of 0.85 and value ratio close to 1.0 when the mean flux density is above 6.0\,Mx\,cm$^{-2}$. 
It can be seen that the helioseismically-derived magnetic-flux maps are not an accurate reflection of the far-side magnetic-flux distributions { when the flux density is small}; however, it is also undeniable that these maps are still useful approximations when neither far-side magnetic fields nor far-side EUV observations are directly available. 

\section{Discussions}
\label{sec5}

Modeling the Sun's global-scale coronal magnetic field and solar wind structures requires synchronous photospheric magnetic field on both the Sun's near and far sides; however, in the foreseeable future, routine observations of the Sun's far-side magnetic fields { will not be }available. 
These facts make the far-side magnetic-flux maps, estimated { by} applying machine-learning methods { to} helioseismic far-side images, useful for global magnetic field users despite their relatively low spatial resolution and accuracy.

In this article, we have demonstrated that it is feasible to use the \euvi's 304\,\AA\ images to bridge the near- and far-side observations.
We have first trained a machine-learning model through pairing the near-side magnetic-field observations and 304\,\AA\ images observed by the \sdo\ instruments, and satisfactorily converted the 304\,\AA\ images into magnetic-flux maps. 
We then apply the trained machine-learning model on recalibrated \stereo-observed 304\,\AA\ images to produce far-side magnetic-flux maps, which are coupled with the far-side helioseismic images for a second machine-learning training. 
The second trained model is thus able to convert the far-side helioseismic images into magnetic-flux maps.
Because the far-side helioseismic images can be obtained using only the near-side continuous Doppler-velocity observations, currently available through both space- and ground-based observations, this means that we are able to, approximately, map the far-side magnetic flux using only the near-side observations.

The resulting far-side magnetic-flux maps are clearly useful to the modeling efforts that require synchronous near- and far-side magnetic fields as inputs. 
Particularly, when a rapid AR emergence { event} occurs on the far side, or when a near-side AR continues to grow substantially after rotating onto the far side, our data products can provide crucial magnetic information of the far side that may change the global-scale magnetic configuration. 
Our estimated magnetic-flux maps, though systematically weaker in peak values, are not significantly off in locations and sizes of the ARs.

However, there are also clear limits in our far-side magnetic-flux estimates. 
First, we can only provide unsigned magnetic-flux density estimates rather than signed magnetic field estimates. 
This is because both the 304\,\AA\ emission and the helioseismic waves' phase shifts are only sensitive to magnetic-field strengths, but insensitive to the polarities of the fields, { making it not straightforward to assign polarities to the estimated magnetic flux.
 Meanwhile, we also recognize that signed magnetic fields were obtained by some authors using deep-learning methods from the \aia\  EUV maps \citep[e.g.,][]{kim19}, and in following studies from \stereo\ EUV maps as well \citep{Jeo20, Als20, Jeo22}. While it is feasible to expand our first machine-learning model to incorporate signed magnetic field with similar neural-networks,  it is not feasible for the second machine-learning model, since the input helioseismic images have a much worse resolution and lack the fine-structure details like those in the EUV maps from which the machine-learning models can extract information.} 
{ A reasonable course of action is to} assign polarities to the magnetic flux based on Hale's law, i.e., assign the leading polarity to the leading part of the AR and following polarity to the following part of the AR. 

{ This approach represents an educated guess based on the well-known statistical properties of AR polarities.}
Another way is to couple the flux transport model \citep[e.g.,][]{upt14} with our magnetic-flux maps, with the former providing magnetic polarity information and the latter providing total magnetic flux. 
However, this method does not work for new emergence on the Sun's far side. 

Another limit in our far-side magnetic-flux maps is that some small-scale { features}, which appear like small ARs, do not actually correspond to true far-side ARs. 
This is because the noise levels are quite high in the helioseismic far-side imaging method. 
As pointed out by \citet{zha19}, { large features are almost always associated with a true far-side AR. Specifically, those features with} areas greater than {  100
Helioseismic Image Pixels (HIPs, about 0$\fdg$80 $\times$ 0$\fdg$83 at disk center) }  have a probability of 97.3\% corresponding to a true far-side AR. 
{ Statistically, such large ARs are very likely to be detected by the helioseismic method. True ARs having a size greater than 200 HIPs on EUV maps were detected helioseismically in 85.7\% of cases. The detection rate rises nearly 100\% when the AR spans at least 400 HIPs.}
 { 
Note that our machine learning model only empirically converts the acoustic signals to unsigned magnetic-flux values, so the sensitivity in the detection of true ARs, and the precision of the detected ARs, are  still limited by the helioseismic detection capability.} 
In this regard, the method developed by \citet{fel19} and \citet{bro21} will likely offer a better guidance on which regions we can have more confidence in and which ones we cannot. 

Despite the limitations on our far-side magnetic-flux maps, we still believe that generating such maps using machine-learning methods is a useful step toward a better modeling of the Sun's far-side magnetic field without sending a spacecraft for routine observations.
The quality of these maps can likely be further improved by further refining the helioseismic far-side imaging technique and, possibly, by further progress of machine-learning algorithms.

\begin{acknowledgments}
Both \sdo\ and \stereo\ are NASA missions. 
\hmi\ is an instrument developed by Stanford University under a NASA contract, and \aia\ is an instrument developed by LMSAL under a NASA contract.
The \euvi\ synchronic maps are generated by the \stereo/SECCHI team and maintained at JHUAPL, in collaboration with NRL and JPL.
This work was partly sponsored by a NOAA grant NA18NWS4680082.
\end{acknowledgments}


\begin{thebibliography}{}

\bibitem[Alshehhi (2020)]{Als20}Alshehhi, R. 2020, in Proceedings of the IEEE/CVF
Conference on Computer Vision and Pattern Recognition Workshops, 204–205

\bibitem[Arge et al.(2010)]{arg10} Arge, C. N., Henney, C. J., Koller, J., Compeau, C. R., Young, S., MacKenzie, D., Fay, A., Harvey, J. W. 2010, AIP Conf. Proc., 1216, 343

\bibitem[Arge et al.(2013)]{arg13} Arge, C. N., Henney, C. J., Gonzalez Hernandez, I., Toussaint, W. A., Koller, J., Godinez, H. C. 2013, AIP Conf. Proc., 1539, 11

\bibitem[Boerner et al.(2014)]{boe14} Boerner, P. F., Testa, P., Warren, H., Weber, M. A., \& Schrijver, C. J. 2014, \solphys, 289, 2377 

\bibitem[Braun \& Lindsey(2001)]{bra01} Braun, D. C., \& Lindsey, C. 2001, \apjl, 560, L189

\bibitem[Broock et al.(2021)]{bro21} Broock, E. G., Felipe, T., Asensio Ramos, A. 2021, \aap, 652, A132

\bibitem[Caplan et al.(2016)]{cap16} Caplan, R. M., Downs, C., \& Linker, J. A. 2016, \apj, 823, 53

\bibitem[Chamberlin et al.(2007)]{cha07} Chamberlin, P. C., Woods, T. N., \& Eparvier, F. G. 2007, Space Weather, 5, S07005

\bibitem[Chamberlin et al.(2008)]{cha08} Chamberlin, P. C., Woods, T. N., \& Eparvier, F. G. 2008, Space Weather, 6, S05001

\bibitem[Felipe \& Asensio Ramos(2019)]{fel19} Felipe, T., Asensio Ramos, A., 2019, \aap, 632, A82 

\bibitem[Fukushima \& Miyake(1982)]{fuk82} Fukushima, K., \& Miyake, S. 1982, Competition and cooperation in neural nets, pp. 267-285, Springer, Berlin, Heidelberg.

\bibitem[Galvez et al.(2019)]{gal19} Galvez, R., Fouhey, D. F., Jin, M., et al. 2019, \apjs, 242, 7

\bibitem[Gonz\'alez Hern\'andez et al.(2007)]{gon07} Gonz\'alez Hern\'andez, I., Hill, F., \& Lindsey, C. 2007, \apj, 669, 1382

\bibitem[Hartlep et al.(2008)]{har08} Hartlep, T., Zhao, J., Mansour, N. N., Kosovichev, A. G. 2008, \apj, 689, 1373

\bibitem[Howard et al.(2008)]{how08} Howard, R. A., Moses, J. D., Vourlidas, A., et al. 2008, \ssr, 136, 67

\bibitem[Ilonidis et al.(2009)]{ilo09} Ilonidis, S., Zhao J., \& Hartlep, T. 2009, \solphys, 258, 181

\bibitem[Jeong et al.(2020)]{Jeo20} Jeong, H.-J., Moon, Y.-J., Park, E., et al.\ 2020, \apjl, 903, L25. 

\bibitem[Jeong et al.(2022)]{Jeo22} Jeong, H.-J., Moon, Y.-J., Park, E., et al.\ 2022, arXiv:2204.12068

\bibitem[Kim et al.(2019)]{kim19} Kim, T., Park, E., Lee, H., et al. 2019, Nature Astron, 3, 397

\bibitem[Kingma \& Ba(2014)]{Adam} Kingma, D.P.  \&  Ba, J., 2014. arXiv preprint arXiv:1412.6980.

\bibitem[Lemen et al.(2012)]{lem12} Lemen, J. R., Title, A. M., Akin, D. J., et al. 2012, \solphys, 275, 17 

\bibitem[LeCun et al.(1989)]{lec89} LeCun, Y., Boser, B., Denker, J., Henderson, D., Howard, R., Hubbard, W., \& Jackel, L. 1989, Advances in neural information processing systems, 2

\bibitem[Liewer et al.(2014)]{lie14} Liewer, P. C., Gonz\'alez Hern\'andez, I., Hall, J. R., Lindsey, C., \& Lin, X. 2014, \solphys, 289, 3617

\bibitem[Liewer et al.(2012)]{lie12} Liewer, P. C., Gonz\'alez Hern\'andez, I., Hall, J. R., Thompson, W. T., \& Misrak, A. 2012, \solphys, 281, 3

\bibitem[Liewer et al.(2017)]{lie17} Liewer, P. C., Qiu, J., \& Lindsey, C. 2017, \solphys, 292, 146

\bibitem[Lindsey \& Braun(2000a)]{lin00} Lindsey, C., \& Braun, D. C. 2000a, Science, 287, 5459

\bibitem[Lindsey \& Braun(2000b)]{lin00b} Lindsey, C., \& Braun, D. C. 2000b, \solphys, 192, 261

\bibitem[Lindsey \& Braun(2017)]{lin17} Lindsey, C., \& Braun, D. C. 2017, Space Weather, 15, 761

\bibitem[Nitta \& DeRosa(2008)]{nit08} Nitta, N. V., \& DeRosa, M. L. 2008, \apjl, 673, L207

\bibitem[Ronneberger et al.(2015)]{ron15} Ronneberger, O., Fischer, P., \& Brox, T. 2015, In: Navab, N., Hornegger, J., Wells, W., Frangi, A. (eds) Medical Image Computing and Computer-Assisted Intervention -- MICCAI 2015. Lecture Notes in Computer Science, vol 9351, 234, Springer, Cham. https://doi.org/10.1007/978-3-319-24574-4\_28 

\bibitem[Scherrer et al.(2012)]{sch12a} Scherrer, P. H., Schou, J., Bush, R. I., et al. 2012, \solphys, 275, 207

\bibitem[Schou et al.(2012)]{sch12b} Schou, J., Scherrer, P. H., Bush, R. I., et al. 2012, \solphys, 275, 229

\bibitem[Schrijver \& DeRosa(2003)]{sch03} Schrijver, C. J., \& DeRosa, M. L. 2003, \solphys, 212, 165

\bibitem[Schrijver \& Title(2011)]{sch11} Schrijver, C. J., \& Title, A. M. 2011, \jgr, 116, A04108

\bibitem[Ugarte-Urra et al.(2015)]{uga15} Ugarte-Urra, I., Upton, L., Warren, H. P., \& Hathaway, D. H. 2015, \apj, 815, 90

\bibitem[Upton \& Hathaway(2014)]{upt14} Upton, L., \& Hathaway, D. H.  2014, \apj, 780, 5

\bibitem[Woods et al.(2012)]{woo12} Woods, T. N., Eparvier, F. G., Hock, R., et al. \solphys, 275, 115

\bibitem[W\"ulser et al.(2004)]{wul04} W\"ulser, J.-P., Lemen, J. R., Tarbell, T. D., et al. 2004, Telescopes and Instrumentation for Solar Astrophysics. ed. Fineschi, S., Gummin, M. A. Proc. SPIE, v.5171, 111

\bibitem[Zhao(2007)]{zha07} Zhao, J. 2007, \apjl, 664, L139

\bibitem[Zhao et al.(2019)]{zha19} Zhao, J., Hing, D., Chen, R., \& Hess Webber, S. 2019, \apj, 887, 216

\end{thebibliography}
\end{document}